\documentclass[twocolumn,english,a4]{revtex4}
\usepackage[T1]{fontenc}
\usepackage[latin9]{inputenc}
\usepackage{amsmath}
\usepackage{amssymb}
\usepackage{graphicx}

\makeatletter

\providecommand{\tabularnewline}{\\}

\@ifundefined{textcolor}{}
{%
 \definecolor{BLACK}{gray}{0}
 \definecolor{WHITE}{gray}{1}
 \definecolor{RED}{rgb}{1,0,0}
 \definecolor{GREEN}{rgb}{0,1,0}
 \definecolor{BLUE}{rgb}{0,0,1}
 \definecolor{CYAN}{cmyk}{1,0,0,0}
 \definecolor{MAGENTA}{cmyk}{0,1,0,0}
 \definecolor{YELLOW}{cmyk}{0,0,1,0}
}

\usepackage{babel}

\makeatother

\usepackage{babel}
\begin{document}

\title{Structure and stability of hydrogenated carbon atom vacancies in
graphene}

\author{Marina Casartelli$^{a}$, Simone Casolo$^{a}$, Gian Franco Tantardini$^{a,b}$and
Rocco Martinazzo$^{a,b}$}

\email{rocco.martinazzo@unimi.it}

\selectlanguage{english}%

\affiliation{$^{a}$ Dipartimento di Chimica, Università degli Studi di Milano,
via Golgi 19, 20133 Milan, Italy.\\
 $^{b}$ Istituto di Scienza e Tecnologia Molecolari, Consiglio Nazionale
delle Ricerche, via Golgi 19, 20133 Milan, Italy. }
\begin{abstract}
Adsorption of hydrogen atoms to a carbon atom vacancy in graphene
is investigated by means of periodic \emph{first principles} calculations,
up to the fully hydrogenated state where six H atoms chemically bind
to the vacancy. Addition of a single H atom is highly exothermic and
barrierless, and binding energies remain substantial for further hydrogenation,
with a preference towards structures with the least number of geminal
pairs. Thermodynamic analysis shows that defective graphene is extremely
sensitive to hydrogenation, with the triply hydrogenated anti- structure
prevailing at room temperature and for a wide range of H$_{2}$ partial
pressures, from $\sim1$ bar down to $<10^{-20}$ bar. This structure
has one unpaired electron and provides a spin-half local magnetic
moment contribution to graphene paramagnetism. Comparison of our results
with recent TEM, STM and $\mu$-SR experiments suggest that carbon
atom vacancies may actually be hydrogenated to various degrees under
varying conditions. 
\end{abstract}
\maketitle

\section{Introduction}

Defects in graphene such as ad-species or missing carbon atoms -commonly
referred to as ``$p_{Z}$ vacancies''- play important roles in charge
transport, magnetism and chemistry of graphene. Most often they are
unavoidable byproducts of the fabrication process (for instance, when
reducing graphene oxide) but they can also be deliberately introduced
by cold plasma treatment, heavy-ion bombardment or high-energy electrons,
in the latter case employing the same transmission-electron-microscopy
(TEM) setup used to image the surface\cite{KrasheninnikovReview10}.
Vacancies and other under-coordinated sites are known to be starting
sites for metal-induced etching\cite{Q-etch,Boggild,Severin}, to
show an enhanced chemical reactivity\cite{bonfanti11,ReviewTerrones}
and to incorporate dopant species\cite{Q-SiEELS}. 

All the above defects introduce semi-localized $\pi$ ``midgap''
states which decay slowly (as $\sim1/r$) from the defect position\cite{pereira08a,Pereira2006,Ugeda10}
and host itinerant electrons which are free to move on one of the
two sublattices of which graphene is made (the one opposite to the
sublattice containing the closest defect). As such, they act as resonant
scatterers and determine graphene conductivity at zero and finite
carrier densities\cite{PeresDisorder,Robinson2008,Wehling2010,PeresRMP,Ferreira2011},
provide spin-half semi-local magnetic moments detectable by magnetometry
experiments\cite{Nair12,Nair13} and bias graphene chemical reactivity
towards specific lattice positions\cite{Hornekaer2006a,casolo09}. 

Carbon atom vacancies occupy a special position in this context since,
in addition to the above $\pi-$''midgap'' state, they have $\sigma$
dangling bonds arising upon atom removal from the breaking of the
C-C bonds which held the lattice atom in place. In the bare vacancy
a structural (Jahn-Teller) distortion occurs that removes two of these
bonds, as predicted by theory\cite{Elbarbary03,Lethinen2004,Yazyev2007,Dharma08,Dai11,palacios12,Casartelli13}
and confirmed by scanning tunneling microscopy (STM)\cite{Mizes1989,Weiss-STM,Takahiro-STMVac,STM-D3hvac,Niimi,Ugeda10}
and TEM\cite{TEM-Robertson} experiments. Even though is not clear
whether such distortion is static or dynamic\cite{Elbarbary03}, thereby
breaking or preserving the threefold symmetry of the ideal vacancy,
it leaves one single $\sigma$ electron which is free to couple with
the above $\pi$ one. As a result, the ground state of a bare vacancy
is a \emph{triplet}, but the singlet with one spin flipped (which
is only $\sim0.2$ eV higher in energy) can be easily accessed if
ripples in the graphene sheet or interaction with a substrate are
taken into account\cite{Casartelli13}. However, the newly formed
C-C bond is rather weak, as expected from its length ($\sim$2.0 \AA{})
which is much larger than a typical single C-C bond, and can in principle
be broken by addition of \emph{e.g.} H atoms. Thus, under typical
laboratory conditions, the question arises to what extent hydrogenation
occurs, and which structures result. This question is of interest
in many respects: the degree of hydrogenation (and their relative
arrangement) determines the residual magnetic moment of the defect,
influences the structure of the carbon atom vacancy (which might be
locked in a distorted geometry\cite{TEM-Robertson} or be statically
symmetric), and possibly explain recent muon spin-resonance ($\mu$-SR)
experiments suggesting formation of geminal adducts of muonion species
on singly hydrogenated vacancies.

In this paper we study the structure, the stability, the magnetic
properties and the mechanism of the carbon mono-vacancy hydrogenation,
from one to six hydrogen atoms, and discuss the relevance of our results
to a number of experimental investigations. The paper is organized
as follows: Section II outlines the method used and the adopted computational
set-up, Section III discusses the results and Section IV summarizes
and concludes.

\section{Theory}

The structure and the energetics of single and multiple hydrogenated
vacancies in graphene was investigated in periodic models with plane-wave
density functional theory (DFT) as implemented in the Vienna \emph{ab
initio} simulation package (VASP)\cite{VASP1,VASP4}. The size of
the supercell, as well as the relevant computational parameters (k-mesh,
plane wave cutoff for expanding the wavefunction and the charge density,
etc.) were selected by preliminary tests similarly to our recent work\cite{Casartelli13}
and chosen to be a good compromise between the need of minimizing
interactions between periodic images and computational cost. The exchange-correlation
effects were included \emph{via} the Perdew-Burke-Ernzerhof \cite{PBE1,PBE2}
functional within the generalized gradient approximation, in its spin-polarized
form. Kohn-Sham orbitals were expanded in a plane-wave basis set limited
to a 500 eV energy cutoff and core electrons described by the projector
augmented-wave\cite{PAW1,PAW2} method. A $6\times6$ supercell with
$20\:\textrm{\AA}$ vacuum between periodic replica normal to the
surface was used, and the Brillouin zone was sampled with a $6\times6\times1$,
$\Gamma$-centered $k$-point mesh; geometry relaxation was performed
till the Hellman-Feyman forces on each atom decreased below 0.01 eV/\AA{}.

Reaction energies for adsorbing $n$ hydrogen atoms were computed
as 
\begin{equation}
\Delta E_{n}=E_{VH_{n}}-E_{V}-E_{nH}\label{eq:formation E}
\end{equation}
where $E_{VH_{n}}$ is the total energy of the $n-$times hydrogenated
vacancy (denoted as $VH_{n}$), $E_{V}$ that of the bare vacancy
(in its ground-state) and $E_{nH}$ is the reference state of $n$
hydrogen atoms, $E_{nH}=nE_{H}$. The latter, along with the molecular
energy $E_{H_{2}}$ to be introduced below, was obtained with a setup
similar to that described above, using a $\Gamma$ point calculation
on a cubic cell with a 20 $\textrm{\AA}$ long side. Reaction energies
for \emph{sequential} adsorption, $\Delta\Delta E_{n}$, were defined
similarly as
\[
\Delta\Delta E_{n}=\Delta E_{n}-\Delta E_{n-1}=E_{VH_{n}}-E_{VH_{n-1}}-E_{H}
\]
(where, unless otherwise stated, $\Delta E_{n-1}$ refers to the most
stable $n-1$-times hydrogenated state) and used to compare the relative
strength of the different CH bonds which may form during the adsorption
process. We also computed a few selected activation energies (barriers)
following the nudged elastic bands method\cite{NEB2} as implemented
in VASP. 
\begin{figure}
\begin{centering}
\includegraphics[clip,width=0.7\columnwidth]{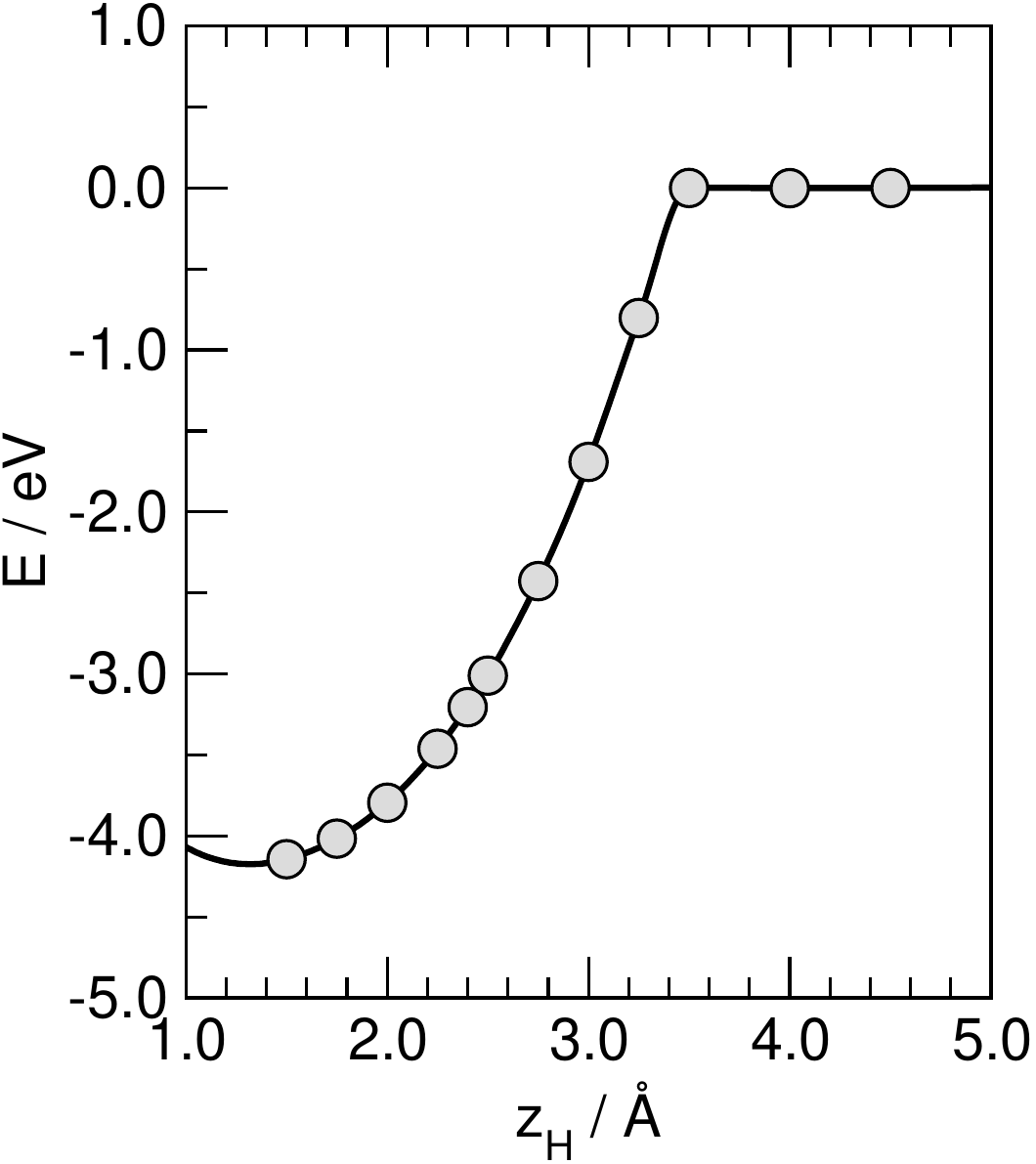}
\par\end{centering}

\caption{\label{fig:Adsorption-energy-profile}Adsorption energy profile for
a single hydrogen atom approaching the vacancy as a function of its
height above the surface. }

\end{figure}

\section{Results}

\subsection{Structure and energetics }

\begin{figure*}[!tp]
\begin{centering}
\includegraphics[clip,width=1\columnwidth]{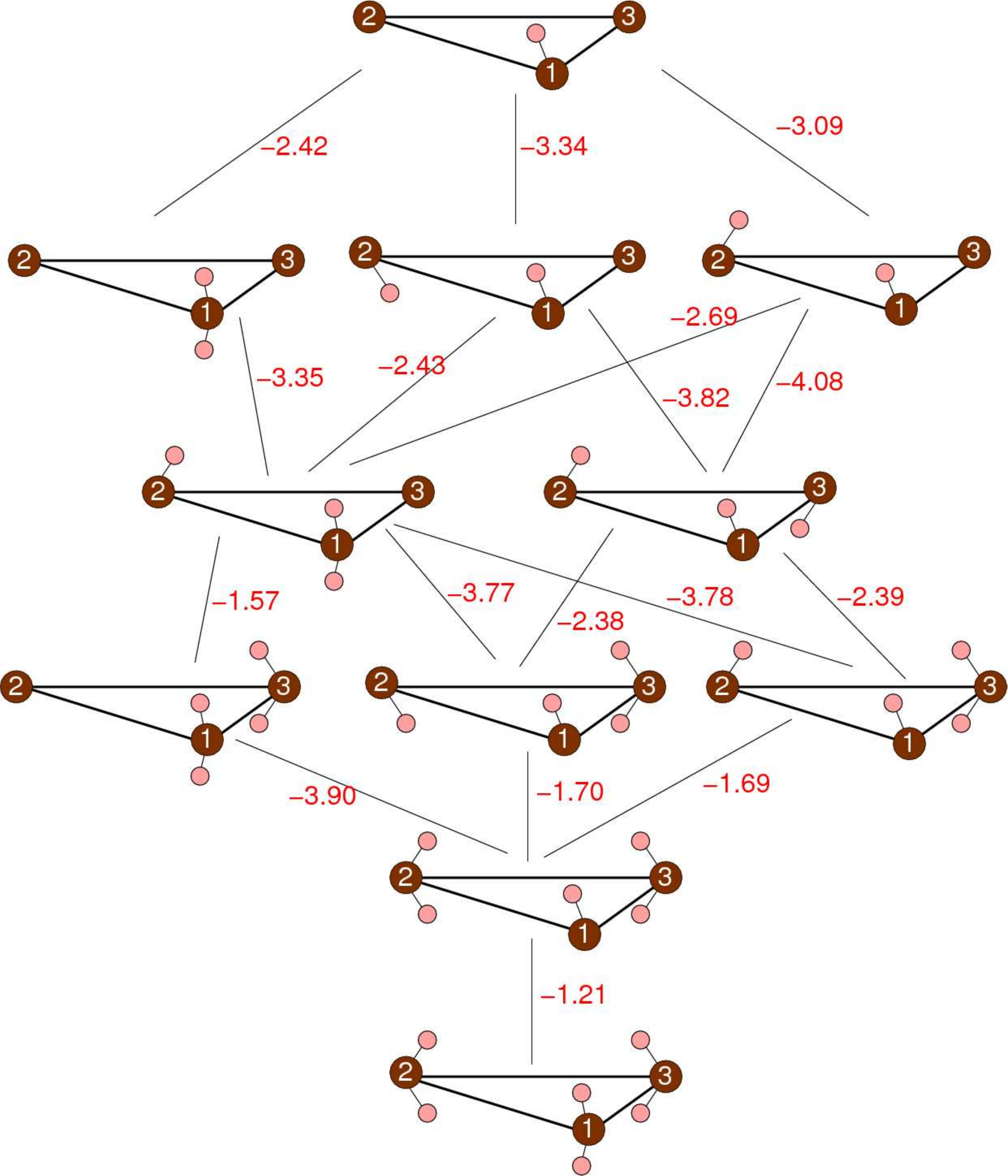}
\par\end{centering}

\caption{\label{fig:Detailed-map-of}Detailed map of the energetics of the
hydrogenated vacancy, showing a schematics of the optimized structures,
with the energies $\Delta\Delta E_{n}$ between them. }

\end{figure*}
As mentioned in the introduction, a vacancy in graphene forms upon
the extraction of a carbon atom which leaves three $\sigma$ dangling
bonds and one unpaired $\pi$ electron on the lattice. The $\sigma$
electrons are tightly localized on the sites which are nearest neighbors
of the vacant one, while the $\pi$ electron delocalizes on one of
the two sublattices (similarly to the case of several monovalent chemisorbed
species such as H and F) as a consequence of the aromatic character
of the substrate\cite{casolo09}. As a result, the bare vacancy is
three-fold symmetric in a doubly-degenerate electronic state and undergoes
a Jahn-Teller (JT) distortion, where two of the three carbon atoms
above close a pentagon by forming a weak CC bond \cite{Elbarbary03,Lethinen2004,Yazyev2007,Dharma08,Dai11,palacios12,Casartelli13},
which may be referred to as JT-induced C-C bond. There are three such
possibilities which differ in the identity of the carbon atom opposing
the pentagon, the so-called \emph{apical} atom. The remaining unpaired
electrons give rise to singlet and triplet manifolds whose relative
stability mainly depends on the height of such apical carbon atom
above the surface\cite{Casartelli13}, the planar triplet being most
stable. Within the present theoretical approach the electronic ground
state has `dirty' magnetization $M=1.56\,\mu_{B}$ because of both
the allowance of fractional occupation and the spin-polarized framework
(see discussion on this point on Ref. \onlinecite{Casartelli13}),
but has an energy which is only few meV below than that of the correct
$M=2\,\mu_{B}$ state. Thus, in the following we use this state as
a reference state for the bare vacancy.

Adsorption of a first H atom is strongly exothermic, $\Delta E_{1}=-4.24$
eV, and barrierless, as shown in Fig.\ref{fig:Adsorption-energy-profile}
where we plot the adsorption profile obtained when optimizing all
the atom coordinates but the height of the hydrogen atom above the
surface. The H atom binds to the apical C atom where much of the unpaired
electron density resides, and a strongly covalent bond forms between
the two (notice for comparison that a typical value for the CH binding
energy is $\sim4$ eV). The carbon atom moves slightly out of the
plane (and correspondingly the CH bond gets tilted), suggesting that
full $sp^{2}-sp^{3}$ re-hybridization of its valence orbitals occurs.
This structure has one unpaired electron left, even though the computed
magnetization ($M=0.56\,\mu_{B}$) is lower than one, for the same
reasons mentioned above for the bare vacancy; indeed, the energy obtained
upon constraining the magnetization to $M=1\,\mu_{B}$ is only few
meV higher. 

When adsorbing a second H atom at the vacancy site a more complex
scenario arises since the two adsorbates may bind to either the same
C atom (\emph{geminal} configuration) or two distinct C sites (\emph{dimer}
configuration). In the latter case, the two atoms may share the same
(\emph{syn-}) or the opposite (\emph{anti-}) graphene face. All these
processes are exothermic when referenced to H atoms {[}much more than
H atom adsorption on the basal plane of graphene/graphite, $\Delta E=-0.97$
eV according to the most recent works\cite{Ivanovskaya2010}{]} but
with a clear preference towards formation of an \emph{anti}-dimer:
$\Delta\Delta E_{2}=-2.42,-3.09$ and $-3.34$ eV for the geminal,
the \emph{syn-} and \emph{anti}-dimer respectively. Activation barriers
are small and probably beyond the accuracy limits of the adopted computational
approach, but suggestive of a fast adsorption kinetics: they are $22$
meV for the geminal product and $21$ meV for both dimers. 

The above doubly-hydrogenated structures have different magnetic properties,
with a net magnetization $M=2.00\,\mu_{B}$ for the \emph{anti-} dimer
and $M=0.00\,\mu_{B}$ for the geminal one (the \emph{syn-} structure
turns out to have $M=0.00\,\mu_{B}$ with \emph{two} unpaired electrons,
and is thus similar in many respects to the \emph{anti-} isomer).
This follows from the different binding processes involved. The first
case corresponds \emph{overall} to the addition of two H atoms on
the \emph{weak} JT C-C bond of the reconstructed vacancy, a process
which breaks such C-C bond but leaves the two unpaired electrons of
the bare vacancy unaltered. On the other hand, the geminal product
may be thought as obtained by double addition to the apical C atom
and corresponding saturation of its unpaired electron density. This
is confirmed by the geometrical parameters of the optimized structures:
the shortest C-C bond of the bare vacancy ($d_{CC}=1.98$ \AA{}; for
comparison the remaining two bonds are $2.72$ \AA{} long) is almost
unaffected in the geminal adduct ($d_{CC}=2.06,2.58,2.58$ \AA{})
while is largely increased (above the value of the longer bonds) in
the dimer structures, $d_{CC}=2.82,2.72,2.72$ \AA{} and $d_{CC}=2.80,2.65,2.65$
\AA{} for the \emph{anti-} and \emph{syn-} structures respectively.
Notice also that, among the latter, the \emph{syn-} structure experiences
an unfavorable H-H steric interaction which forces the system to additional
relaxation and reduces its stability compared to the \emph{anti-}
one by $\sim23$ meV. 

Next, we consider adsorption of further hydrogen atoms to the three
carbon atoms surrounding the vacancy, and their relative arrangements,
see the schematics reported in Fig. \ref{fig:Detailed-map-of} as
a reference. Similarly to the sticking of the second H atom, three
hydrogen atoms may have a geminal pair or sit on different sites.
In the latter case they are considered as \emph{syn-} when they \emph{all}
sit on the same face of the graphene sheet, and \emph{anti-} otherwise.
With four H atoms formation of a geminal pair is unavoidable, yet
the arrangement of the remaining two atoms may be geminal or \emph{syn-}/\emph{anti-}
dimer. The five and six atom cases parallel in a complementary way
the single H and bare vacancy cases with just one possible arrangement. 

The main results of this study can be summarized as follows. Among
the three-atom adducts the \emph{anti}- structure is $\sim1.4$ eV
most stable than the geminal one ($\Delta\Delta E_{3}=-3.82$ eV \emph{vs.}
$\Delta\Delta E_{3}=-2.43$ eV) and the \emph{syn}- structure is not
binding at all. The two structures have one unpaired electron (we
obtained $M=1.00\,\mu_{B}$ in both cases) either because the additional
H atom breaks the JT C-C bond and leaves one unpaired electrons (if
we start from the doubly hydrogenated geminal structure and form the
VH$_{3}$ geminal one) or because it couples (at low spin) with one
of the two unpaired electrons of the \emph{anti}- dimer structure
and form either the geminal or the \emph{anti}- VH$_{3}$ structure.
In any case, the JT C-C bond no longer exists and the C-C distances
are considerably increased ($d_{CC}=2.75,2.72,2.62$ \AA{} in the
geminal and $d_{CC}=2.86,2.85,2.81$ \AA{} in the \emph{anti}- structure).
The energetics follows consistently: starting from the most stable
\emph{anti}- dimer structure addition of a third H atom is easier
(and more exothermic) when a $\sigma$ bond is formed on the apical
atom than when $H$ binds to an already hydrogenated site to form
a geminal product, since in the latter case a bond is formed with
the $\pi$-like electron left unpaired on that site. 

Among the possible four-fold hydrogenated structures the \emph{syn}-
is slightly more stable than the \emph{anti}- ($\Delta\Delta E_{4}=-2.39$
eV \emph{vs. }$\Delta\Delta E_{3}=-2.38$ eV) and definitely more
stable than the geminal one ($\Delta\Delta E_{4}=-1.57$ eV starting
from the geminal VH$_{3}$ structure). At this stage all the four
unpaired electrons initially available upon vacancy formation get
saturated with hydrogen, and both the \emph{syn}- and the \emph{anti}-
structures are non-magnetic, differently from the geminal one which
has two unpaired electrons (we obtained $M=0.00\,\mu_{B}$ in the
first two cases and $M=2.00\,\mu_{B}$ in the latter). Further hydrogenation
is remarkably less exothermic ($\Delta\Delta E_{5}=-1.69$ eV and
$\Delta\Delta E_{6}=-1.21$ eV) since it necessarily requires further
breaking of the aromaticity, and produces magnetic structures with
one and two unpaired electrons, respectively. Indeed, compared to
the bare vacancy, full hydrogenation requires (i) saturation of the
three $\sigma$ dangling bonds, (ii) saturation of the $\pi$ (midgap)
state and (iii) breaking of \emph{two} C-C double bonds (a first C-C
double bond needs to be broken when hydrogenating the VH$_{4}$ \emph{anti}-
or \emph{syn}- structure).
\begin{figure}
\begin{centering}
\includegraphics[clip,width=0.8\columnwidth]{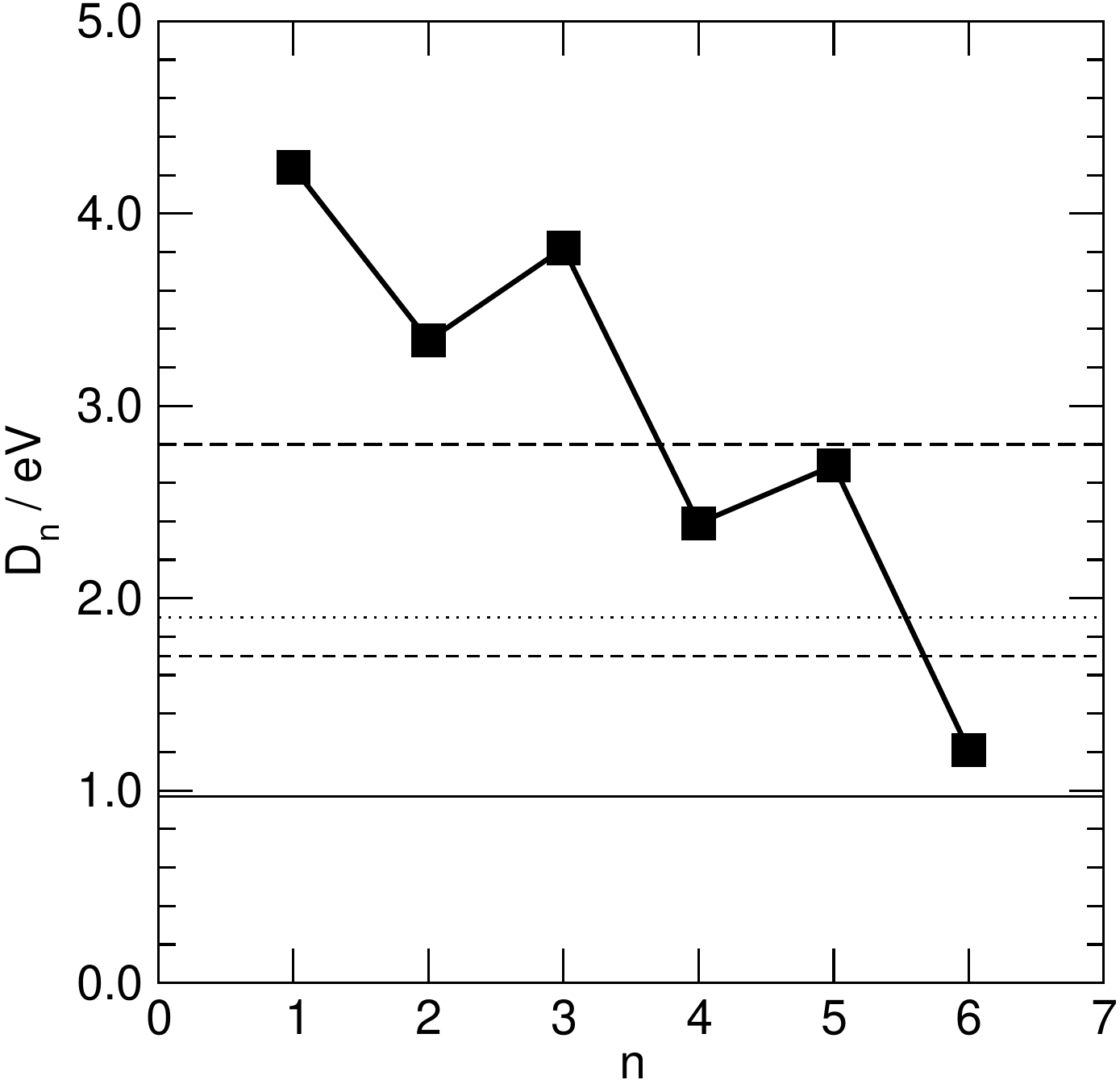}
\par\end{centering}

\caption{\label{fig:H-atom-binding}H atom binding energies for the most stable
$VH_{n}$ structures. Horizontal lines are reference values for the
H binding energy in the `bulk' (full line), for the secondary H adsorption
to form  \emph{para-} dimers (dotted line), and for adsorption at
an armchair (dashed) or zig-zag (long-dashed) edge.}
\end{figure}

Fig. \ref{fig:H-atom-binding} gives an overview of the energetics
of the hydrogenated carbon atom vacancy in graphene. There, the H-atom
\emph{binding} \emph{energies }(or \emph{affinities}) $D_{n}=-\Delta\Delta E_{n}$
to form the most stable $n$-times hydrogenated structure from the
most stable $n-1$ isomer are plotted as a function of $n$. Such
structures are the \emph{anti}- structures, the $n=4$ case being
an exception which shows a slight ($\sim0.01$ eV) preference towards
the \emph{syn}- structure. For comparison, the values of the H binding
energy in several different situations are given as horizontal lines;
such values comprise the adsorption energy to the basal graphene plane\cite{Ivanovskaya2010},
the binding energy of the secondary H atom to form a \emph{para}-
dimer\cite{casolo09} (a similar result holds for the \emph{orto-
}dimer\cite{casolo09}) and the binding energy for adsorption on an
armchair edge and a zig-zag edge which are H terminated\cite{Ahlrichs00}.
The latter represents the largest energy gain when adsorbing H on
pristine graphene and compare rather well with the binding energies
of the fourth and fifth hydrogen atoms; larger energies are obtained
for the first three H atoms where bonding involves $\sigma$ orbitals
and/or unpaired electrons. 

Of interest are also some general structural changes accompanying
hydrogenation. We find that structures with $n\ge4$ are more open
than the less hydrogenated ones, with $d_{CC}\gtrsim2.7$ \AA{}. This
value has to be compared with the corresponding value in graphene
(\emph{i.e.} the next-to-nearest neighbors C-C distance in the bare,
unrelaxed vacancy) which is $d_{CC}=2.46$ \AA{}. This `expansion'
is accompanied by a substantial movement of the C atoms out of the
plane, up to $0.7-0.8$ \AA{} . For instance, in the VH$_{3}$ \emph{anti}-
structure we obtained $\delta z_{C}=0.68,049$ and $-0.34$ \AA{}
and in the VH$_{5}$ structure $\delta z_{C}=0.79,0.66$ and $0.18$
\AA{}. C-H bond lengths fall in the range $1.07-1.12$ \AA{}, bonds
in geminal pairs being the longest and almost similar to CH in perfect
graphene ($1.13$ \AA{}). With the same token, H-H distances in geminal
pairs all fall in the range $d_{HH}=1.73-1.79$ \AA{} and distances
in \emph{syn}- and \emph{anti}- pairs are generally larger than these
values, but smaller values are possible for \emph{syn}- pairs. For
instance we found $d_{HH}=1.51$ \AA{} for the VH$_{2}$ \emph{syn}-
structure and $d_{HH}=1.42$ \AA{} for the \emph{syn}- pair in the
VH$_{3}$ \emph{anti}- structure.

\noindent .

\subsection{Thermodynamic analysis}

\begin{figure}
\begin{centering}
\includegraphics[clip,width=0.8\columnwidth]{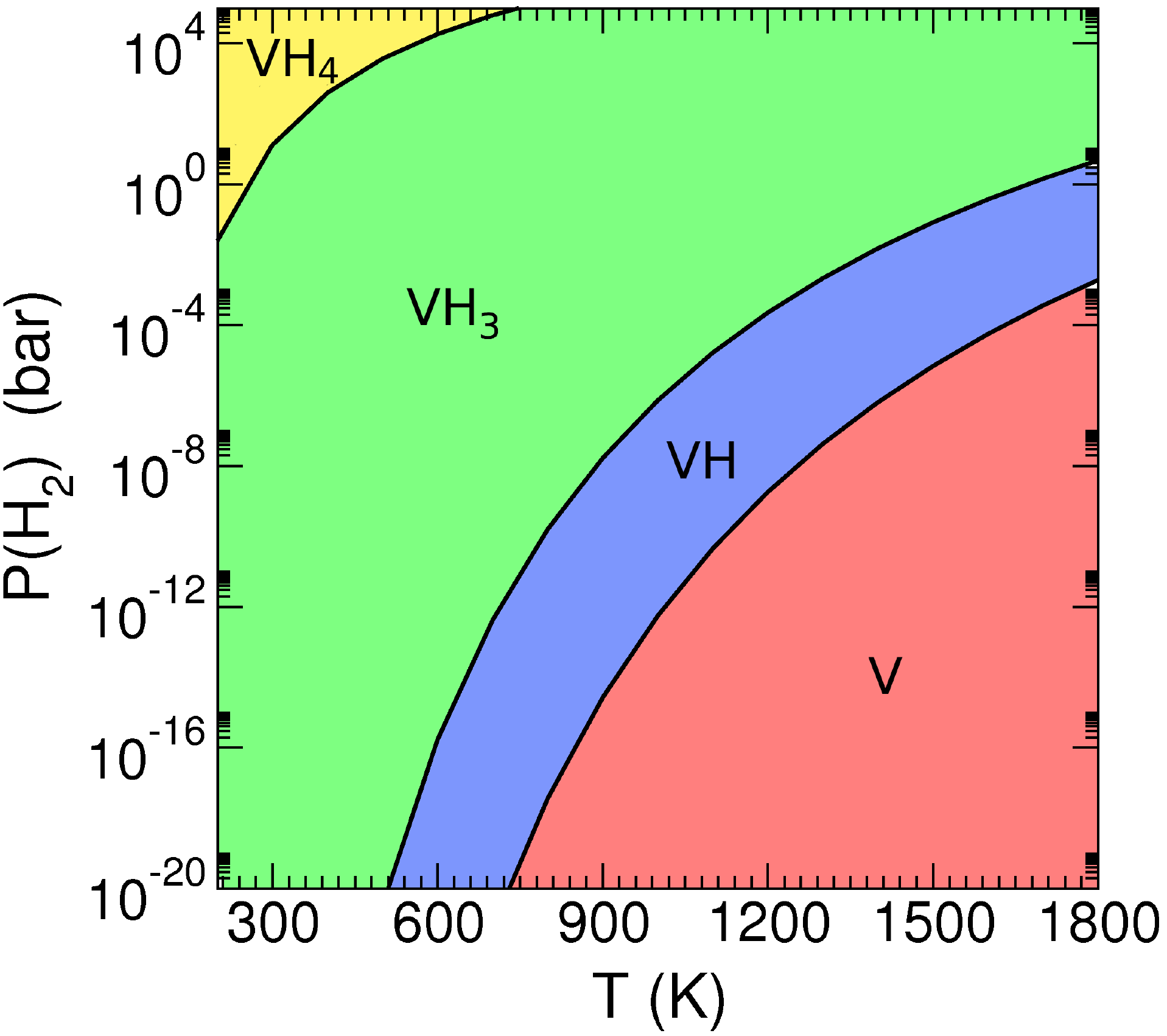} 
\par\end{centering}

\caption{\label{fig:Phase-diagram-for}Phase diagram for the hydrogenated C-vacancy
in graphene, showing the most stable hydrogenated species $VH_{n}$. }
\end{figure}
With the whole energetics of the hydrogenated vacancy at hand we performed
a thermodynamic analysis to investigate the relative stability of
the various hydrogenated species under reasonable hydrogen partial-pressure
and temperature conditions. We considered the lowest molar Gibbs free
energy of formation $\Delta G_{n}^{f}(p,T)$ at a H$_{2}$ partial
pressure $p$ and temperature $T$, which -analogously to Eq.(\ref{eq:formation E})
- is given by
\[
\Delta G_{n}^{f}(p,T)=G_{VH_{n}}(p,T)-G_{V}(p,T)-\frac{n}{2}G_{H_{2}}(p,T)
\]
where $G_{VH_{n}}$, $G_{V}$ and $G_{H_{2}}$ are Gibbs free energies
for the $n-$times hydrogenated vacancy, the bare vacancy and molecular
hydrogen, respectively. For the first two terms we neglected any temperature
and pressure dependence and thus relied on the DFT energies above
($G_{VH_{n}}\simeq E_{VH_{n}}$ for $n=0,1,$ etc.). For hydrogen,
some thermodynamic data are available from standard tables\cite{webbook}
and ideal-gas dependence on its partial pressure $p$ can be assumed
to write 
\begin{align*}
G_{H_{2}}(p,T) & =H_{H_{2}}(p^{\ominus},T)-TS_{H_{2}}(p^{\ominus},T)+\\
 & +RT\ln\left(\frac{p}{p^{\ominus}}\right)\equiv E_{H_{2}}+\Delta G_{H_{2}}(p,T)
\end{align*}
where $R$, as usual, is the perfect gas constant, $S_{H_{2}}(p^{\ominus},T)$
is the hydrogen molar entropy at temperature $T$ and standard pressure
$p^{\ominus}=1$ bar, and the enthalpy is written as $H_{H_{2}}=E_{H_{2}}+\frac{5}{2}RT$
for $T<T^{\ominus}$ and $H_{H_{2}}=E_{H_{2}}+\frac{5}{2}RT^{\ominus}+\Delta H_{H_{2}}(p^{\ominus},T)$
otherwise. Here $E_{H_{2}}$ is the DFT energy of H$_{2}$ in its
equilibrium configuration and $\Delta H_{H_{2}}(p^{\ominus},T)$ is
the H$_{2}$ molar enthalpy change from $T=T^{\ominus}=298.15$ K
to $T$ at the standard pressure $p^{\ominus}$(Ref. \onlinecite{webbook}).
$\Delta G_{H_{2}}(p,T)$ is defined by the above equation and is the
appropriate free-energy change from $T=0$ to $T,\, p$. Accordingly,
\[
\Delta\Delta G_{n}=\Delta G_{n}^{f}-\Delta G_{n-1}^{f}=\Delta\Delta E_{n}^{m}-\frac{1}{2}\Delta G_{H_{2}}(p,T)
\]
where $\Delta\Delta E_{n}^{m}$ differs from the energy for sequential
adsorption defined above by half the dissociation energy $D_{m}$
of the $H_{2}$ molecule, 
\[
\Delta\Delta E_{n}^{m}=E_{VH_{n}}-E_{VH_{n-1}}-\frac{1}{2}E_{H_{2}}=\Delta\Delta E_{n}+\frac{1}{2}D_{m}
\]
where $D_{m}=4.5$ eV with our setup.  

The results of such an analysis are reported in Fig. \ref{fig:Phase-diagram-for}
in the temperature range $T=250-1800$ K and for a wide hydrogen partial
pressure range $p=10^{-20}-10^{5}$ bar, comprising standard atmospheric
conditions ($T\sim300\, K$ and $p=5.55\times10^{-7}$ bar in dry
atmosphere) and ultra high vacuum conditions (UHV) where the pressure
is $\sim10^{-12}$ times smaller than the standard value; $p=1$ bar
and higher, on the other hand, can be achieved upon hydrogen exposure.
Fig. \ref{fig:Phase-diagram-for} shows that the bare vacancy is thermodynamically
stable only at high temperature and low hydrogen pressure; for instance,
thermal annealing at $T>1200$ K would be required in atmospheric
conditions to free vacancy defects from hydrogen atoms, whereas annealing
at $T=600\, K$, while sufficient for desorbing H atoms and dimers
from the basal plane\cite{Hornekaer2006}, is not enough to desorb
H atoms form the vacancy. Even under UHV conditions annealing at $T\gtrsim800$
K is required to have bare vacancies, thereby suggesting that, at
least for naturally occurring vacancies, hydrogen passivation is ubiquitary
and extensive, with the triply hydrogenated species most abundant
up to $T\sim600$ K. Different results may be expected when vacancies
are intentionally introduced into the substrate by \emph{e.g.} electron
or ion bombardment, since the hydrogenation kinetics and ensuing equilibration
may be considerably slowed down by the low operating partial pressures
and temperatures. For instance, STM experiments are often performed
at low temperatures ($T\sim10$ K) where a VH$_{3}$ isomer is the
most stable species under a wide range of pressures, but where the
hydrogenation kinetics is extremely slow unless high hydrogen partial
pressures are used. Similarly for TEM experiments, despite the larger
temperature typically used (ambient conditions).
\begin{figure}
\begin{centering}
\includegraphics[width=0.8\columnwidth]{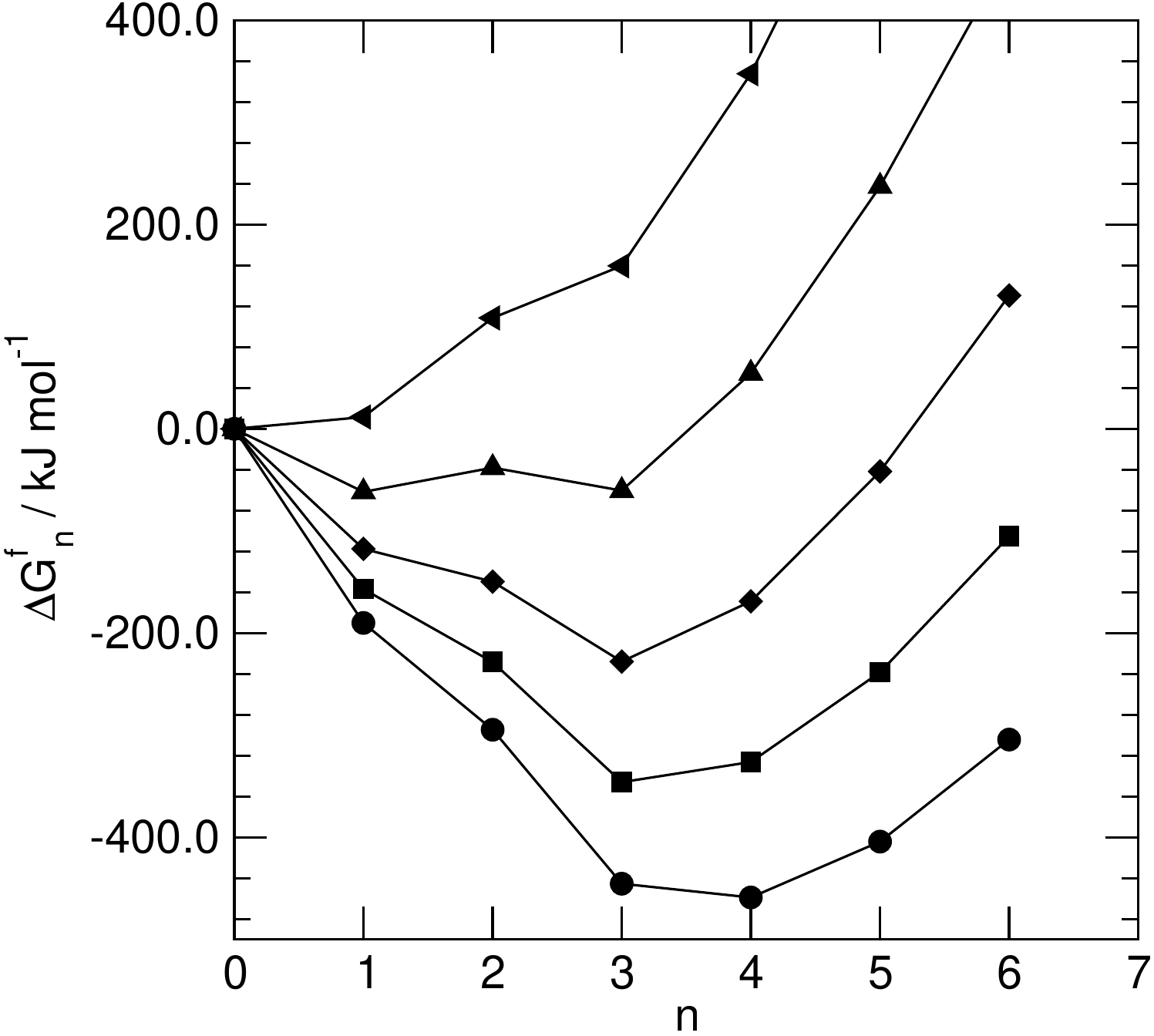}
\par\end{centering}

\caption{\label{fig:Molar-free-energy-of}Molar free-energy of formation for
the most stable isomers $VH_{n}$ at atmospheric conditions $p=0.55\times10^{-6}$
bar, for different temperatures. Circles, squares, rhombus, triangles
up and triangles down for $T=0,300,600,1000$ and $1500$ K, respectively.}
\end{figure}

Noteworthy is the absence in Fig. \ref{fig:Phase-diagram-for} of
any doubly hydrogenated species, neither the geminal nor the \emph{syn}-/\emph{anti}-
dimer: upon increasing the temperature at constant pressure the most
stable isomers change from the triply hydrogenated form to the singly
hydrogenated one, without passing from a ($p,T$) region where dimers
are stable against hydrogenation/dehydrogenation; see also Fig. \ref{fig:Molar-free-energy-of}
which reports the molar free-energies of the most stable isomers for
different temperatures, at ambient hydrogen partial pressure. 

Finally, it is worth noticing that, as a consequence of the energetics
discussed in the previous subsection, no geminal species prevails
at equilibrium. Geminal adducts have been recently invoked to explain
long-time oscillations in the zero-field $\mu$-SR decay asymmetry\cite{RiccoNL,Ricco13},
which were attributed to dipolar interactions between a muonion (Mu,
a H isotope $\sim$10 times lighter than H) and a H atom in a -CHMu
geminal structure. Though not stable at the experimental conditions
($T=300$ K) these structures might easily form from the stable singly-
and triply- hydrogenated species prevailing under a wide range of
conditions, and dominate the $\mu$-SR signal over alternative arrangements
because of structural reasons. Among these, as observed above, the
H-H distance in geminals has a well defined value ($d_{HH}\simeq1.75$
\AA{}, in good agreement with the dipolar distance extracted by the
$\mu$-SR signal, $d=1.70\pm0.02$ \AA{}), irrespective of the specific
structure and unaffected by low-frequency vibrations of the lattice
(the lowest-frequency motion altering such distance is a H-C-H bending).
However, further investigation is required to establish whether such
a product can indeed form in a collision between a muonion and a hydrogenated
vacancy.

\subsection{Magnetic properties }

The magnetic properties of carbon atom vacancies in graphene have
been recently subjected to experimental investigations with conflicting
results. Early report of paramagnetism due to spin-1 species\cite{ney11},
in accordance with a triplet ground-state configuration of the bare
vacancy\cite{Casartelli13}, have been questioned on the light of
magnetometry measurements on carefully prepared samples, which showed
the presence of spin-$1/2$ species for both flourine ad-species and
vacancies\cite{Nair12}, at odds with the different numbers of unpaired
electrons left upon defect formation. Later measurements from the
same group showed that in the presence of carbon atom vacancies \emph{two}
different spin-$1/2$ local magnetic moments contribute to the observed
paramagnetic signal, one arising from the $\sigma$ network and one
from the $\pi$ band system\cite{Nair13}. This finding would agree
with the above mentioned electronic structure of the Jahn-Teller distorted
vacancy if the singlet-triplet separation (Hund's coupling) were vanishing
small to allow a decoupled response of the two unpaired electrons
left upon reconstruction, a possibility which has been recently ruled
out by accurate \emph{ab} \emph{initio} results\cite{Casartelli13}.
In the same work it has been shown that the energetics is sensitive
to the out-of-plane motion of the apical carbon atom, thereby suggesting
that ripples and/or interactions with a substrates could explain why
the above mentioned unpaired electrons of the vacancy are uncoupled.
Here we consider a different scenario, \emph{i.e.} the possibility
that the vacancy is hydrogenated to some extent and thus presents
a net magnetic moment which depends on the degree of hydrogenation.
\begin{table}
\noindent \begin{centering}
\begin{tabular}{|l|c|c|c|}
\hline 
 & $M/\mu_{B}$  & $m$ & Character\tabularnewline
\hline 
\hline 
1H  & 0.56  & 1 & $\pi$\tabularnewline
\hline 
2H geminal  & 0.00  & 0 & -\tabularnewline
\hline 
2H anti  & 2.00  & 2 & $\sigma$, $\pi$\tabularnewline
\hline 
2H syn  & 0.00  & 2 & $\sigma$, $\pi$\tabularnewline
\hline 
3H geminal  & 1.00  & 1 & $\sigma$\tabularnewline
\hline 
3H anti  & 1.00  & 1 & $\pi$\tabularnewline
\hline 
4H geminal  & 1.92  & 2 & $\sigma$, $\pi$\tabularnewline
\hline 
4H anti  & 0.00  & 0 & -\tabularnewline
\hline 
4H syn  & 0.00  & 0 & -\tabularnewline
\hline 
5H  & 1.00  & 1 & $\pi$\tabularnewline
\hline 
6H  & 2.00  & 2 & $\pi$, $\pi$\tabularnewline
\hline 
\end{tabular}
\par\end{centering}

\centering{}\caption{Computed total magnetization $M$ in $\mu_{B}$ and predicted number
$m$ of unpaired electrons left on the C-vacancy after the hydrogenation
process, along with their character.\label{tab:mag}}
\end{table}

Table \ref{tab:mag} summarizes the computed magnetic moment for each
structure investigated, along with the number of unpaired electrons
predicted with the help of the resonating valence bond model, and
their `character',\emph{ i.e.} whether they are $\sigma$ or $\pi$
and thus localized on a single lattice position or delocalized over
several lattice sites. Notice though that ``$\pi$-moments'' entries
in the Table are most often hybrid $\sigma-\pi$ moments: for instance,
the singly hydrogenated structure can be thought as originated from
binding a H atom to one of the two $sp^{3}$ hybrid orbitals of the
apical carbon atom of the bare vacancy, a process which leaves one
unpaired electron in a $sp^{3}$ orbital free to hybridize with $p_{Z}$
states of neighboring C atoms and form $sp^{3}-\pi$ states. 

With only few exceptions the computed magnetic moments refer to pure
spin states and thus agree well with the number of unpaired electrons
predicted by the resonance structures. For instance -as already mentioned
above- among the doubly hydrogenated structures, the \emph{geminal}
one has no magnetic moment since it results from saturating \emph{both}
dangling bonds on the apical C atom of the bare vacancy, while the
\emph{anti-} structure has the same magnetic moment as the bare vacancy
because the two H atoms bind to the JT bond; interestingly, in \emph{syn-
}isomer, the singlet-triplet ordering seems to be reversed and a net
zero megnetic moment with two unpaired electrons signals an \emph{open-shell}
singlet. 

A quick look at Table \ref{tab:mag} reveals that hydrogenated vacancies
have magnetic moments in the range $0.0-2.0\,\mu_{B}$, even though
the corresponding singly occupied states may have very different character.
For instance, despite the presence of one unpaired electron in both
the VH$_{3}$-geminal and the VH$_{3}$\emph{-anti} structure, the
corresponding singly occupied state has either a $\sigma$ or a $\pi$
origin. More interestingly, taking into account the results of the
previous section, we observe that in a wide range of temperature and
hydrogen partial pressure of interest in many practical situations,
the magnetic moment of the most stable hydrogenated species (VH$_{1}$,
VH$_{3}$ and VH$_{4}$ structures) has one and the same value, namely
$M=1.00\,\mu_{B}$. Thus, with the exception of the bare vacancies,
thermodynamically stable vacant species on a wide range of ambient
conditions are expected to behave as spin-$1/2$ magnetic moments
as a consequence of hydrogenation. 
\begin{figure*}
\includegraphics[width=0.5\columnwidth]{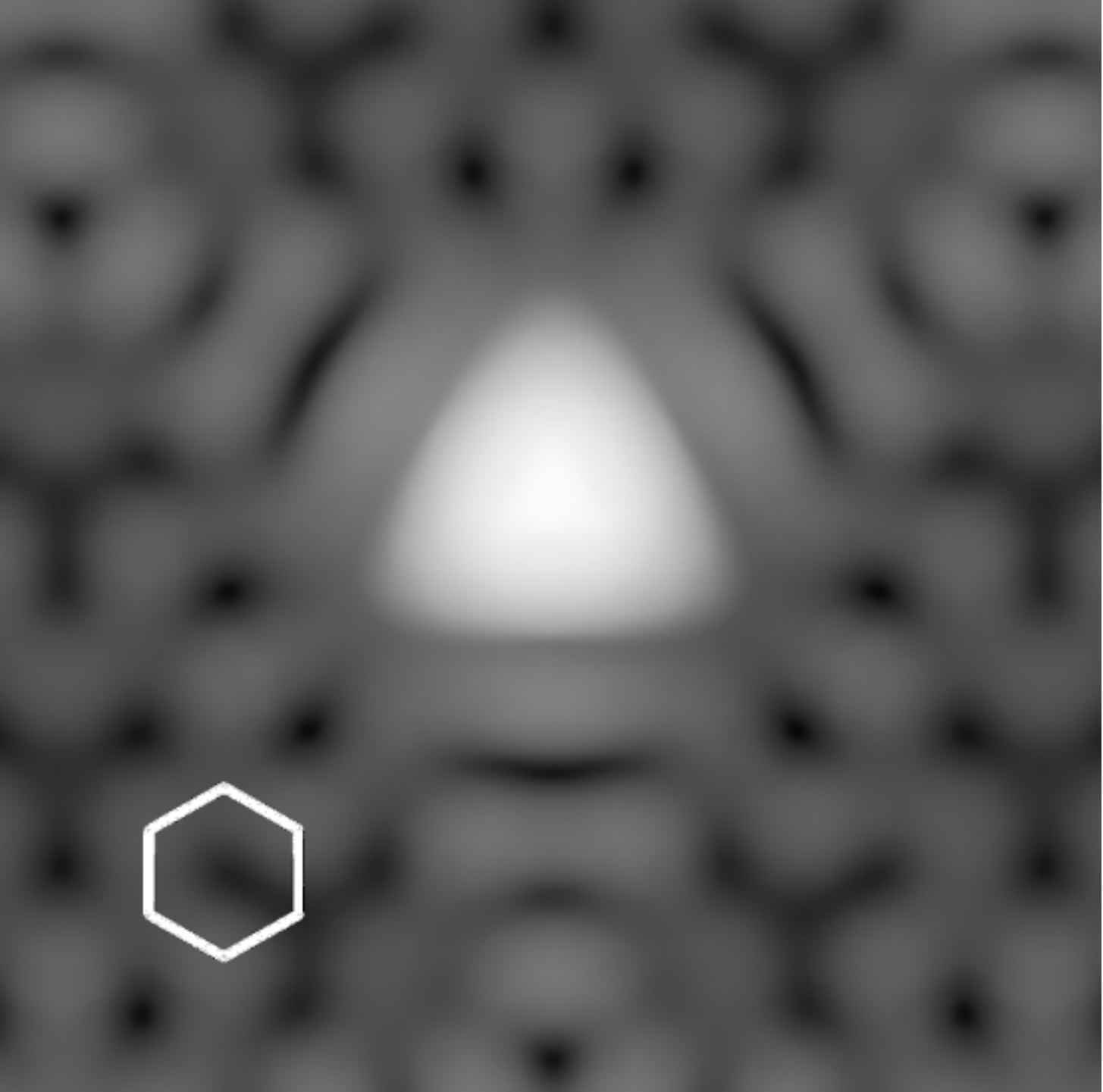}\hspace{0.1cm}(a)
\hspace{0.1cm}\includegraphics[width=0.5\columnwidth]{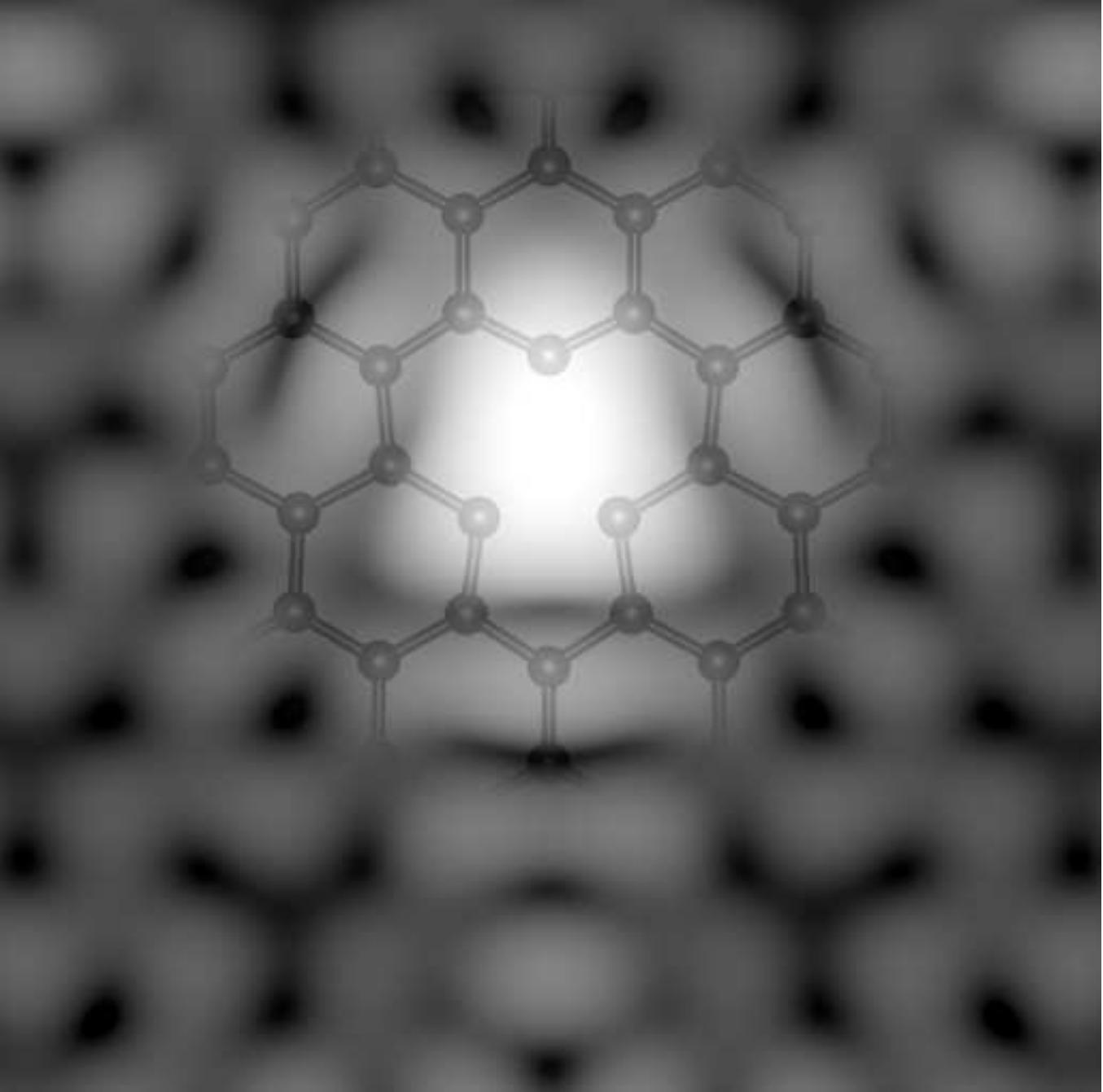}\hspace{0.1cm}(b)\hspace{0.1cm}\includegraphics[width=0.5\columnwidth]{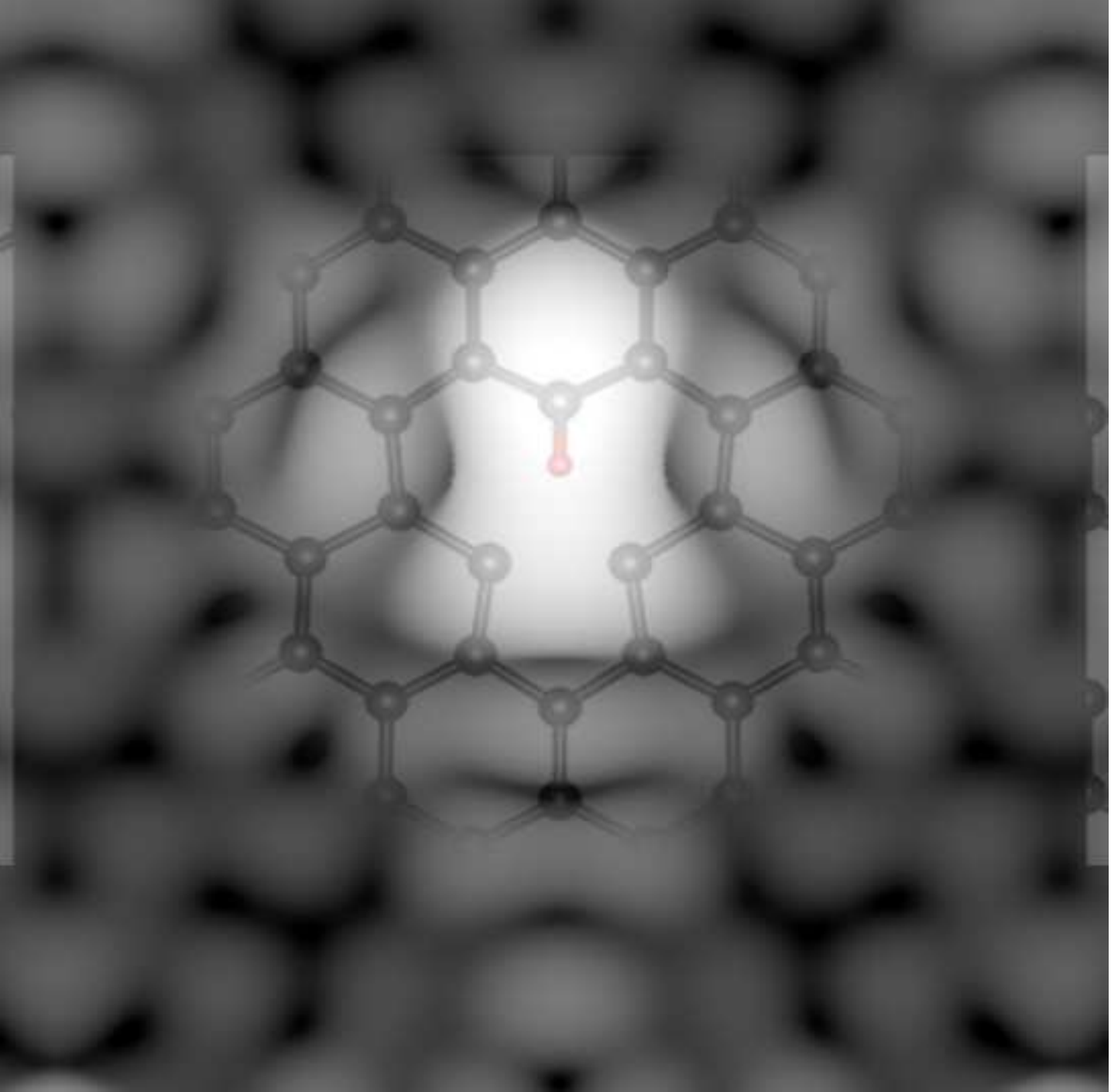}\hspace{0.1cm}(c)

\caption{\label{fig:Simulated-STM-images}Simulated STM images of the vacancy
of the bare vacancy (a,b) and of the singly hydrogenated vacancy (c)
with molecular models showing the lattice arrangement. Image on panel
(a) has been obtained by averaging over the three equivalent orientations
of the structure. }
\end{figure*}

\subsection{STM imaging}

Finally, in order to help identifying and distinguishing the various
hydrogenated species investigated we performed, on a few relevant
optimized structures, simulations of the STM images within the standard
Tersoff-Hamann approximation. The sample bias was fixed at $-0.5\, V$
and constant-current images obtained by integrating the local density
of states in the range $[-0.5,0]$ eV from the Fermi level, as the
(largest) height $z$ of an isosurface of the resulting function. 

Several situations were envisaged. First we considered the bare vacancy
which, in its stable structure, should present a distorted configuration,
with a C-C distance sensibly shorter than the other two. However,
since three equivalent minima are possible, each differing by the
orientation of the JT bond, we also considered the possibility that
orientational averaging occurs on the time-scale of imaging (few seconds).
This could happen either because of thermal hopping between the minima,
or because a dynamic Jahn-Teller effect is operative whereby the ground-state
structure is a superposition of the three possible distorted one and
robust against environmental dephasing. Fig. \ref{fig:Simulated-STM-images}
shows these two possibilities, panel (a) for the averaged structure
and panel (b) for the distorted one, the latter with a superimposed
molecular model to help identifying the lattice positions. As is evident
from Fig.s \ref{fig:Simulated-STM-images}(a,b), despite the structural
differences, the distorted and the symmetrical configurations of the
bare vacancy are hardly distinguishable from each other. Both structures
present a bright signal with exact or approximate three-fold symmetry
which extends far from the vacant site and which corresponds to the
above mentioned $\pi$-midgap state, in agreement with STM/STS experiments
\cite{Ugeda10}, but no further distinguishing feature. 

A distorted structure similar to that of the bare vacancy but with
a locked configuration appears upon single hydrogenation. In this
case, the attached H atom is not free to move on neighboring C atoms
of the vacancy (even though we did not compute it, the barrier for
hopping likely matches the desorption barrier, as happens for hydrogen
atoms adsorbed on the basal plane of graphene\cite{Hornekaer2006a,casolo09})
and the simulated STM image, reported in Fig. \ref{fig:Simulated-STM-images}
(c), shows a clearer breaking of symmetry than in the case of the
bare vacancy. It further shows an increase of brightness in the vacancy
at the expense of the a reduced intensity of the signal due to the
$\pi$- midgap state, probably related to the admixture of $\sigma$
and $\pi$ character (remember that the H atom is accommodated out-of-plane
with a CH bond originating from a $sp^{3}$ C orbital, and an unpaired
electron is left on the remaining $sp^{3}$ orbital, free to hybridize
with the $\pi$-band states. Fig. \ref{fig:Simulated-STM-images}
is the simulated image for the structure with the H atom above the
plane). 

\begin{figure}
\begin{centering}
\includegraphics[width=0.6\columnwidth]{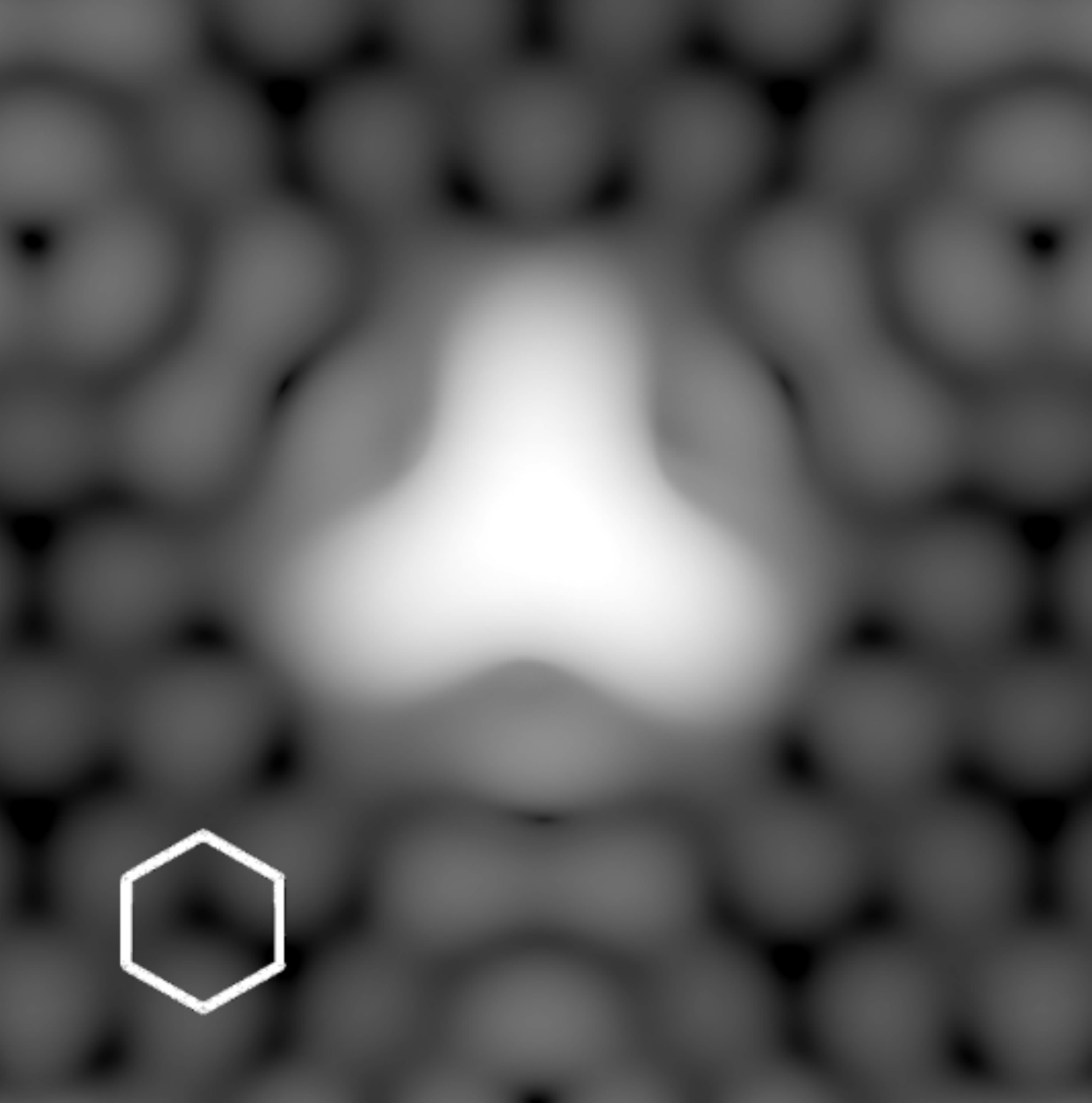}
\par\end{centering}

\caption{\label{fig:Simulated-STM-image 2}Simulated STM images of the most
stable triply hydrogenated isomer of the vacancy, averaged over its
equivalent configurations. }
\end{figure}

Finally, we considered one further thermodynamically stable structure,
the triply hydrogenated \emph{anti-} structure which prevails under
standard conditions. Here, too, a number of equivalent configurations
is possible (three possible orientations of the dimers in \emph{syn-}
relationships times two possible positions, above or below the surface
plane) and we decided to average over them since their interconversion
does not involve any H atom transfer and should thus be possible under
ordinary temperature conditions. Fig. \ref{fig:Simulated-STM-image 2}
reports the simulated image of the structure which shows a clear increase
of the intensity in the vacant site, as well as the expansion of the
lattice accompanying hydrogenation which was mentioned above. Note
that this structure still bears one unpaired $\pi$-moment of the
kind similar to the one in Fig.s \ref{fig:Simulated-STM-images} (a-c)
and thus, far from the vacant site, displays the same `typical' three-fold
symmetric signal, though with a reduced intensity.

A symmetric structure has been recently observed by Robertson and
co-workers\cite{TEM-Robertson} with TEM, under conditions where the
vacancy would be triply hydrogenated if equilibrium conditions prevailed.
However, our computed C-C distances in the VH$_{3}$-\emph{anti} structure
($d{}_{CC}\sim2.8\,$ \AA{}) are sensibly larger than those found
in the experiment ($d{}_{CC}=2.5\pm0.1$ \AA{}) that the observed
structure is likely to be a bare vacancy which dynamically switches
between its equivalent configurations (or an excited state of the
vacancy with electronic different configuration). 

\section{Conclusions}

We reported on a thorough investigation of the structure and energetics
of several hydrogenated states of a carbon atom vacancy in graphene.
We found large H-atom adsorption energies for the first few hydrogen
atoms (related to saturation of $\sigma$-dangling bonds) which progressively
decrease towards values similar to those found for adsorption at graphene
edges. As a consequence, under equilibrium conditions, triply- and
singly- hydrogenated species prevail over a wide range of hydrogen
partial pressure and temperatures relevant in many experimental situations.
The computed energetics shows that geminal pairs are invariably less
stable than different hydrogen arrangements with adsorbed H atom on
different sites (if possible), preferably in an \emph{anti-} configuration.
Thus, whether geminal products are formed (as suggested by recent
$\mu$-SR experiments\cite{RiccoNL,Ricco13}) requires further investigation
of the \emph{dynamical} aspect of the process. 

Most of the investigated structures have a net non-zero magnetic moment
capable of a paramagnetic signal. In particular, one spin-$1/2$ moment
related to an unpaired electron in a semilocalized $\pi$-midgap state
appears both in the singly- and triply- hydrogenated species which
dominate the $(p,T)$ phase-diagram describing hydrogenation. 

Comparison with existing STM investigations of defective graphene\cite{Ugeda10}
showed agreement for the presence of the signal due to the $\pi$-midgap
state, but further studies are needed to establish whether the observed
vacancies are Jahn-Teller distorted, dynamically averaged or hydrogenated
to some extent. Simulated STM images hardly distinguish a locked from
an averaged bare vacancy but present a few marked features accompanying
hydrogenation which may help identifying hydrogenated species.

\section{Acknowledgments}

The authors acknowledge Quentin Ramasse for useful discussions. This
work has been supported by Regione Lombardia and CILEA consortium
through an ISCRA and LISA (Laboratory for Interdisciplinary Advanced
Simulation) 2012 initiative grants.

\bibliographystyle{apsrev}

\end{document}